# BeCAPTCHA: Behavioral Bot Detection using Touchscreen and Mobile Sensors benchmarked on HuMIdb

*Alejandro Acien, Aythami Morales, Julian Fierrez, Ruben Vera-Rodriguez, Oscar Delgado-Mohatar*

*School of Engineering, Universidad Autonoma de Madrid, Spain*

{alejandro.acien, aythami.morales, julian.fierrez, ruben.vera, oscar.delgado}@uam.es

**Abstract**

In this paper we study the suitability of a new generation of CAPTCHA methods based on smartphone interactions. The heterogeneous flow of data generated during the interaction with the smartphones can be used to model human behavior when interacting with the technology and improve bot detection algorithms. For this, we propose BeCAPTCHA, a CAPTCHA method based on the analysis of the touchscreen information obtained during a single drag and drop task in combination with the accelerometer data. The goal of BeCAPTCHA is to determine whether the drag and drop task was realized by a human or a bot. We evaluate the method by generating fake samples synthesized with Generative Adversarial Neural Networks and handcrafted methods. Our results suggest the potential of mobile sensors to characterize the human behavior and develop a new generation of CAPTCHAs. The experiments are evaluated with HuMIdb[1] (Human Mobile Interaction database), a novel multimodal mobile database that comprises 14 mobile sensors acquired from 600 users. HuMIdb is freely available to the research community.

---

[1] https://github.com/BiDAlab/HuMIdb

## 1. Introduction

The research interest in smartphone devices has been constantly growing in the last years. The capacity of these devices to acquire, process, and storage a wide range of heterogeneous data offers many possibilities and research lines (e.g. user authentication [1][2][3][4], health monitoring [5][6][7], behavior monitoring [8][9][10][11], etc). Besides, the usage of mobile phones is ubiquitous. According to [12], mobile lines exceeded world population in 2018, and the amount of smartphones devices sold surpassed world population in 2014 [13]. This is one of the fastest growing manmade phenomena ever, from 0 to 7.2 billion in barely three decades. In the same way, this widget has changed the way we access and create contents on the internet. Recent surveys reveal that nearly three quarters (72.6%) of internet users will access the web via their smartphones by 2025. In fact, almost 51% of web accesses are actually made through mobile phones [14].

On the other hand, mobile web hazards are growing very fast as well. Malicious malware is also adapting to this new mobile era. Mobile bots employ the capacities of smartphones affecting multiples types of online services, such us: social media (e.g. mobile bots accounts propagate fake twitter messages [15]), ticketing/travel, e-commerce, finance, gambling, ATO/Fraud, DDoS attacks, and price scrapping, among others. According to [16], these mobile bots use cellular networks by connecting through cellular gateways. Mobile bots can perform highly advanced attacks while remaining hidden in plain sight. In addition, they are very unlikely to be detected by IP address blocking. [16] showed that 5.8% of all mobile devices on cellular networks are used in malicious bot attacks. In other study [17], researchers reveal that mobile fraud reached 150 million global attacks in the first half of 2018 with attack rates rising 24% year-over-year.

In this context, new countermeasures against fraud adapted to mobile scenarios are necessary. One of the most popular methods to distinguish between humans and bots is known as CAPTCHA (Completely Automated Public Turing test to tell Computers and Humans Apart). These algorithms determine whether the user is human by presenting challenges associated to the cognitive capacities of the human beings. The most common challenges are: recognizing characters from a distorted image (text-based

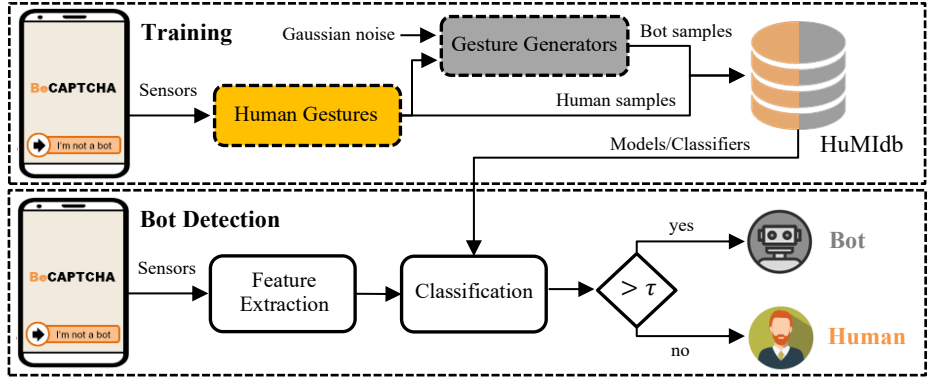

Figure 1: Block diagram of our proposed bot detection system. The response of the bot detector is a combination of responses from two different modalities: touch and accelerometer. $\tau$ is a decision threshold.

CAPTCHAs); identifying class-objects in a set of images (image-based CAPTCHAs); speech translation from distorted audios (sound-based audio CAPTCHAs); or newer systems that replace traditional cognitive tasks by a transparent algorithm capable of detecting bots and humans from their web behavior [18]. However, recent advances in areas such as computer vision, speech recognition, or natural language processing have increased the vulnerabilities of CAPTCHA systems [19][20][21]. Major advances in deep learning applied in those areas enable the generation of synthetic data of very natural appearance, therefore increasingly difficult to detect if used by bots.

Most of the current CAPTCHAs have been designed to be used in a web interaction based on mouse and keyboard interfaces. In this paper we explore the potential of mobile devices to model human-machine interaction for bot detection applications. In particular, we focus here on building a CAPTCHA system (called BeCAPTCHA) based on swipe gestures (i.e. drag and drop task). We model this gesture according to features obtained from the touchscreen and accelerometer sensors in order to extract cognitive and neuromotor human features that help us to discriminate between bots and human users just with simple drag and drop gestures (see Fig.1 for details). To evaluate the CAPTCHA, we employ human samples and synthetic ones (bot-like samples) generated using two different generators: a handcrafted synthesis method and a Generative Adversarial Networks (GANs) generation method. We assume a challenging scenario where the attacker (malicious bot developer) can generate synthetic gestures trying to mimic the sensor signals derived from human-mobile interaction. The goal is to determine whether a simple swipe gesture has been performed by a human or generated by a bot. The main contributions of this work are as follows:

*i)* Summary of relevant recent works in touchscreen biometrics and accelerometer signals for modelling and exploiting the interaction in smartphones.

*ii)* A new method to generate synthetic swipe gestures using GANs and samples acquired during real human-device interaction. This method allows to generate synthetic samples that mimic the human behavior.

*iii)* A new bot detection approach based on modelling the user behavior in smartphone interaction using multiple inbuilt sensors: BeCAPTCHA. We also experiment with a particular implementation of the proposed approach by combining touch dynamics and accelerometer data from HuMIdb, acquired when the users perform swipe gestures. This is a very common gesture used in many touch interfaces (e.g. unlock devices, confirm will to advance to other step).

*iv)* Discussion of relevant CAPTCHA approaches in comparison with the proposed BeCAPTCHA.

*v)* The new public HuMIdb[1] dataset (Human Mobile Interaction database) that characterizes the interaction of 600 users according to 14 sensors during normal human-mobile interactions in an unsupervised scenario with more than 300 different devices.

A preliminary version of this article was presented in [22]. This article significantly improves [22] in the following aspects:

*i)* We significantly augment the positioning with respect to related works.

*ii)* We improve the bot detection accuracy training with real and synthetic samples generated with the adversarial method proposed.

*iii)* We consider a larger number of classifiers and provide an ablation study, exposing the strengths and weakness of the classifiers in each scenario evaluated.

*iv)* We provide a qualitative comparison with traditional CAPTCHA methods and their complementarity with the new method.

*v)* We provide an analysis of user perception about CAPTCHAs technologies: issues and ethics concerns.

The rest of the paper is organized as follows: Section 2 analyzes the capacity of touchscreen, accelerometer and gyroscope mobile sensors for modeling human-machine interaction and summarizes the main existing mobile databases incorporating touchscreen and accelerometer data. In section 3 we introduce HuMIdb, a new multimodal mobile database collected for this work that comprises 14 mobile sensors acquired from 600 users. Section 4 describes the proposed BeCAPTCHA architecture. Section 5 analyzes the results obtained. Section 6 makes a comparison with traditional CAPTCHA methods and the suitability of the proposed BeCAPTCHA to complement the existing ones. Finally, section 7 summarizes the conclusions and future work.

## 2. Mobile Sensors for Modeling Human-Machine Interaction: Accelerometer and Touchscreen

Accelerometer, gyroscope, gravity sensor, touchscreen gestures, keystrokes, light sensor, WiFi, Bluetooth, camera, and microphone are some examples of sensors/signals acquired by a smartphone while we interact with it or just carry it with us during our daily routines. Those data can be used to model human-machine interaction and human behavior. In this section we present examples of different research fields that exploit accelerometer, gyroscope and touch signals obtained or derived from mobile sensors.

- **Accelerometer and gyroscope** are both useful to measure the movements that the smartphone is exposed to: the accelerometer measures the magnitude and direction of acceleration forces applied over the mobile device and the gyroscope measures orientation. These sensors have been studied for mobile user authentication with good results in the last years [23]. For example, in [24] they used these mobile sensors for user recognition through simple gestures like answering a call in four different user states: standing, sitting, walking, and running. In other example, Gafurov *et al.* [25] extracted gait patterns from a mobile device attached to the lower part of the leg in three directions: vertical, forward-backward, and sideways motion. They achieved error rates between 5% and 9% for gait authentication combining all three acceleration measures. Accelerometer has been also studied to measure the daily physical activities with the main goal of changing people's sedentary lifestyle [26]. Mobile apps employing accelerometer and gyroscope to measure physical activity are broadly used among runners, athletes, and healthy people, resulting in a very profitable market. In other research field, accelerometer and gyroscope has demonstrated to be a promising tool for Parkinson disease estimation, identifying Parkinson disease (PD) through physical activities (e.g. walking, standing, sitting, holding) [5] or hand tremor [27]. El-Zayat *et al.* [28] measured the extent of shoulder rotation using the gyroscope. In a similar way, researchers from Korea used the gyroscope sensor to measure the range of motion of the shoulder in subjects suffering from unilateral symptomatic shoulder. They found that this sensor shows an acceptable reliability and high correlation with manual goniometer readings [29].

- **Touchscreen gestures** involve all kind of finger movements that we perform over the smartphone screen (e.g. swipe, tap, zoom). These signals have been studied for user mobile authentication in the last years [1][2]. Nevertheless, it has been shown not to have enough discriminative power to replace traditional authentication technologies such as passwords or swipe patterns, but they achieve good performance in combination with other mobile biometric traits [30][31]. In other research fields, a recent study [32] demonstrates how touch gestures can discriminate between children and adults just with swipe and tap gestures achieving error rates under 5%. That work suggests that touch gestures are ruled by the neuromotor cortex, less developed in children. Touchscreen patterns also provide the possibility to measure aspects of cognitive function. As an example, Apple Research Kit includes tools for standard cognitive tests used in clinical research adapted for smartphone devices such as the spatial memory test, the paced auditory/visual serial addition test (PVSAT), and the simple reaction time test [33]. In other example, the Project EVO app is a mobile game that it is currently being tested in a wide range of clinical studies and patient populations including ADHD, autism, depression, traumatic brain injury, and, more recently, as a biomarker to assess Alzheimer in clinical trials [6].

In summary, the literature demonstrates the potential of mobile sensors to model inner human features including cognitive functions, neuromotor skills, and human behaviors/routines.

| Ref. | Sensors | #Users | Sessions/user | Supervised | Public | # Devices | Task |
|---|---|---|---|---|---|---|---|
| Mahbub *et al.* (2016) [34] | 13 (<u>Tou, Acc</u>, Blu, Cam, Gyr, GPS, Key, Lig, Mag, Press, Prox, Temp, WIFI) | 54 | ~248 | No | Yes | 1 | Free |
| Liu *et al.* (2018) [36] | 5 (<u>Tou, Acc</u>, Gyr, Mag, Pow) | 10 | 3 Hours | Yes | No | 1 | Free |
| Tolosana *et al.* (2019) [37] | 3 (<u>Touch, Acc</u>, Gyr) | 217 | ≤ 6 Sessions | No | Yes | <217 | Fixed |
| **HuMIdb (Present Paper)** | **14 (<u>Tou, Acc</u>, Blu, Gra, GPS, Gyr, Key, LAc, Lig, Mag, Mic, Ori, Prox, WIFI)** | **600** | **≤ 5 Sessions** | **No** | **Yes** | **600** | **Fixed** |

Table 1: Summary of existing mobile databases incorporating at the same time Touchscreen (Tou) and Accelerometer (Acc) signals. Other sensors: Bluetooth (Blu), Front camera (Cam), Gravity (Gra), Gyroscope (Gyr), GPS, Keystroke (Key), Light sensor (Lig), Linear Accelerometer (LAc), Magnetometer (Mag), Microphone (Mic), Orientation (Ori), Power consumption

### 2.1 Mobile Datasets with Touchscreen and Accelerometer Data

Table 1 summarizes previous multimodal mobile databases that include at the same time accelerometer and touchscreen signals and compares them with the new HuMIdb dataset introduced in the present paper.

UMDAA-02 [34] is a multimodal mobile database that includes 14 mobile sensors: front camera, touchscreen, gyroscope, accelerometer, magnetometer, light sensor, GPS, Bluetooth, WiFi, proximity sensor, temperature sensor, and pressure sensor. The data was collected during 2 months from 48 volunteers in an unsupervised scenario with 248 sessions per user in average and using the same smartphone (Nexus 5). In other work [36], the authors collect touch gestures, power consumption, accelerometer, gyroscope, and magnetometer mobile signals from 10 participants under laboratory conditions and with the same mobile device (supervised scenario) during a period of three hours. In [37] the authors collected a database of mobile touch on-line data named MobileTouchDB. The database is focused on mobile touch patterns and contains more than 64K on-line character samples performed by 217 users with a total of 6 sessions. They also acquired accelerometer and gyroscope signals under unsupervised conditions.

## 3. The HuMIdb Database

In this section we introduce the Human Mobile Interaction database (HuMIdb), a novel multimodal mobile database that comprises more than 5 GB from a wide range of mobile sensors acquired under unsupervised scenario. The database includes 14 sensors (see Table 2 for the details) during natural human-mobile interaction performed by 600 users. For the acquisition, we implemented an Android application that collects the sensor signals while the users complete 8 simple tasks with their own smartphones and without any supervision whatsoever (i.e., the users could be standing, sitting, walking, indoors, outdoors, at daytime or night, etc.) The acquisition app was launched on Google Play Store and advertised in our research web site and various research mailing lists. After that, the participants were self-selected around the globe producing more varied participants than previous state-of-the-art mobile databases. All data captured in this database have been stored in private servers and anonymized with previous participant consent according to the GDPR (General Data Protection Regulation).

The different tasks are designed to reflect the most common interaction with mobile devices: keystroke (name, surname, and a pre-defined sentence), tap (press a sequence of buttons), swipe (up and down directions), air movements (circle and cross gestures in the air), handwriting (digits from 0 to 9), and voice (record the sentence '*I am not a robot*'). Additionally, there is a drag and drop button between tasks (see Appendix A for details).

The acquisition protocol comprises 5 sessions with at least 1-day gap among them. It is important to highlight that in all sessions, the 1-day gap refers to the minimum time between one user finishes a session and the next time the app allows to have the next session. At the beginning of each task, the app shows a brief pop-up message explaining the procedure to complete each task. The application also captures the orientation (landscape/portrait) of the smartphone, the screen size, resolution, the model of the device, and the date when the session was captured.

| Sensors | Sampling Rate | Features | Power Consumption |
|---|---|---|---|
| Accelerometer | 200 Hz | $x, y, z$ | Low |
| L.Accelerometer | 200 Hz | $x, y, z$ | Low |
| Gyroscope | 200 Hz | $x, y, z$ | Low |
| Magnetometer | 200 Hz | $x, y, z$ | Low |
| Orientation | NA | $l$ or $p$ | Low |
| Proximity | NA | $cm$ | Low |
| Gravity | NA | $m/s^2$ | Low |
| Light | NA | $lux$ | Low |
| TouchScreen | E | $x, y, p$ | Medium |
| Keystroke | E | $key, p$ | Medium |
| GPS | NA | Lat., Lon., Alt., Bearing, Accuracy | Medium |
| WiFi | NA | SSID, Level, Info, Channel, Frequency | High |
| Bluetooth | NA | SSID, MAC | Medium |
| Microphone | 8 KHz | Audio | High |

Table 2: Description of all sensor signals captured in HuMIdb. E=Event-based acquisition. The timestamp parameter is captured for all sensors.

Regarding the age distribution, 25.6% of the users were younger than 20 years old, 49.4% are between 20 and 30 years old, 19.2% between 30 and 50 years old, and the remaining 5.8% are older than 50 years old. Regarding the gender, 66.5% of the participants were males, 32.8% females, and 0.7% others. Participants performed the tasks from 14 different countries (52.2%/47.0%/0.8% are European, American, and Asian respectively) using 179 different devices.

Fig. 2 shows an example of the handwriting task (for digit "5") and the information collected during the task. Note how a simple task can generate a heterogeneous flow of information related with the user behavior: the way the user holds the device, the power and velocity of the gesture, the place, etc.

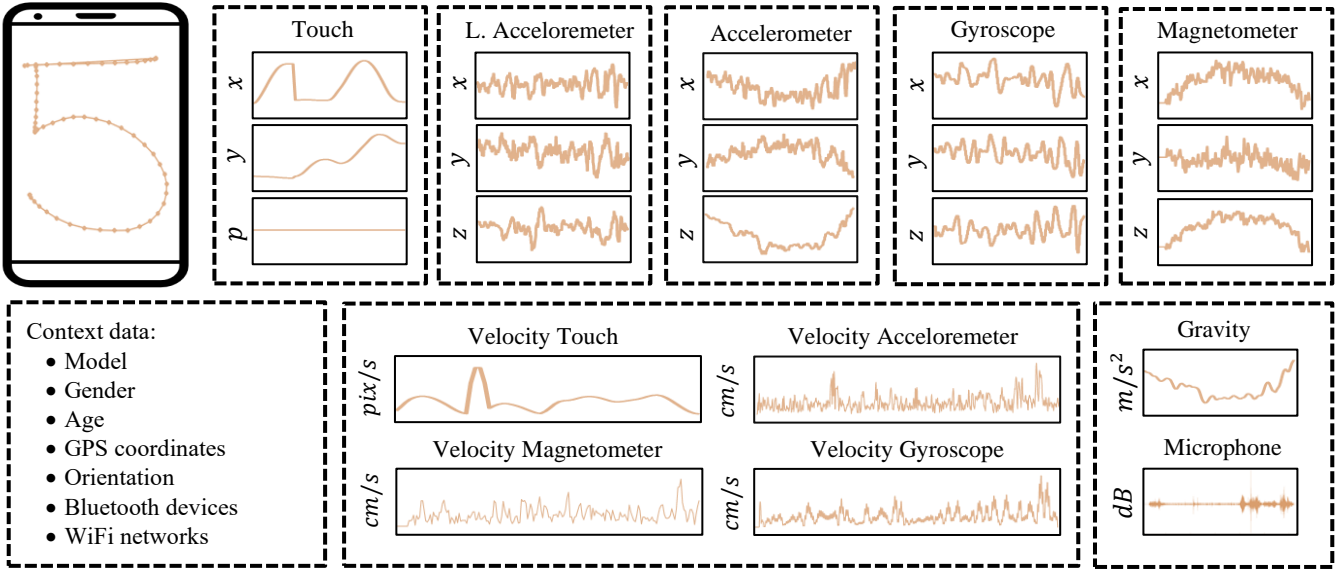

Figure 2: Full set of data generated during one of the HuMIdb task.

## 3.1 HuMIdb Research Opportunities

In this paper we explore the potential of HuMIdb for bot detection, but the richness in number of sensors acquired and population diversity offer many other research possibilities. Some of the possible research lines to explore with this dataset include:

- Demographic modeling: HuMIdb comprises users from the 4 continents and 12 different countries. The database is diverse in gender and age of the participants.
- Cross-sensor interoperability: HuMIdb includes signals from 140 different devices. Analyzing the impact of different device characteristics on human behavior is a challenging research line.
- User recognition: HuMIdb comprises behavioral patterns from 600 users. Continuous authentication based on biometric behavioral patterns is a popular research line with applications in the security market. See for example [38] that could be extended to do continuous authentication.

## 4. BeCAPTCHA: Methods and Experimental Protocol

HuMIdb offers the opportunity to model human behavior. Among the multiple applications, in this work we explore the use of human interaction to develop a new generation of CAPTCHA systems based on mobile inbuilt sensors. In this section we describe the methods and experimental protocol followed for developing BeCAPTCHA, a CAPTCHA system based on swipe gestures plus accelerometer signals (i.e. a drag and drop action when the user scrolls the Next button to the right in HuMIdb). First, we describe how to model this gesture according to features obtained from the touchscreen and accelerometer in order to extract cognitive human features that help us to discriminate between bots and human users (Sect. 4.1). To evaluate the CAPTCHA, we will employ human samples (from HuMIdb) and synthetic ones (bot-like samples) generated using two different approaches: a handcrafted synthesis and using GANs (Sect. 4.2). The goal is to determine whether a simple swipe gesture has been performed by a human or generated by a bot. For this, the experimental protocol is described in Sect. 4.3.

### 4.1 Feature Extraction: Characterizing Swipe Gestures

To characterize swipe gestures from the touchscreen and accelerometer signals, we have adapted two feature sets previously employed in [35][39] for bot detection and user authentication respectively.

The interaction of the user with the Touchscreen is defined by a time sequence $\mathbf{s}_T = \{\mathbf{x}, \mathbf{y}, \mathbf{p}, \mathbf{t}\}$ with length $N$, composed by the coordinates $\{\mathbf{x}, \mathbf{y}\}$, the pressures $\mathbf{p}$ (when available), and the timestamps $\mathbf{t}$. First, the coordinates $\{\mathbf{x}, \mathbf{y}\}$ are normalized by the size of the screen. Second, the pressure is discarded as it is not available in most of the devices. Third, six global features are generated according to Table 3.

The Accelerometer signal is defined by a sequence $\mathbf{s}_A = \{\mathbf{x}, \mathbf{y}, \mathbf{z}, \mathbf{t}\}$. The feature set chosen for the accelerometer signal was adapted from [35], in which they calculate the mean, median, root-mean-square, and standard deviation of the three accelerometer axes $\{\mathbf{x}, \mathbf{y}, \mathbf{z}\}$ for user authentication.

| Parameters | Description |
|---|---|
| Duration ($D$) | $t_{N-1} - t_0$ |
| Distance ($L$) | $\|(x_{N-1}, y_{N-1}) - (x_0, y_0)\|$ |
| Displacement ($P$) | $\sum_{i=0}^{N-1} \|(x_{i+1}, y_{i+1}) - (x_i, y_i)\|$ |
| Angle ($\alpha$) | $\tan^{-1}(|(y_{N-1} - y_0)|/|(x_{N-1} - x_0)|)$ |
| Mean velocity ($V$) | $\frac{1}{N}\sum_{i=0}^{N-1} \|(x_{i+1}, y_{i+1}) - (x_i, y_i)\|/(t_{i+1} - t_i)$ |
| Move Efficiency ($E$) | $P/L$ |

Table 3: Touch features extracted for the characterization of the gestures.

## 4.2 Generating Human-Like Gestures: Bot Samples

A swipe gesture can be defined by a spatial trajectory (sequence of points $\{\mathbf{x}, \mathbf{y}\}$) and a velocity profile determined by the timestamp sequence $\mathbf{t}$. To generate synthetic swipe patterns, we will follow two approaches: handcrafted synthesis and Generative Adversarial Network (GAN) synthesis.

### 4.2.1 Method 1: Handcrafted Synthesis

We observed that most of the human swipe trajectories obtained from our drag and drop task are linear. The handcrafted approach generates swipe trajectories according to a straight-line shape and a realistic velocity profile. For this, we first estimate the probability distribution of length and angle of human swipe gestures in HuMIdb. Note that the size and coordinates of each human swipe varies depending on the device features so we have normalized each one by the total size of the screen.

The synthetic trajectories are defined by the initial point $(x_0, y_0)$, duration ($t_{N-1} - t_0$), angle ($\alpha$), and the velocity profile $\{\mathbf{v}, \mathbf{t}\}$. We have synthesized the fake trajectories according to distributions of these parameters fitted from human data (except for the velocity profile). With the aim to emulate human behaviors, we spaced the points of the linear trajectory on a log scale (emulating a velocity profile with the initial acceleration observed in human samples).

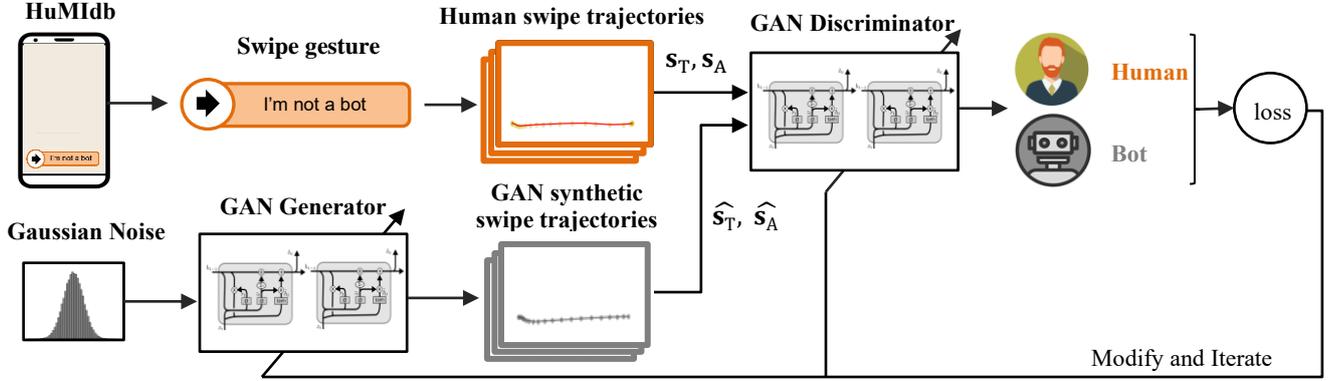

Figure 3: The proposed architecture to train a GAN Generator of synthetic swipe gestures characterized by touch $\widehat{s_T}$ and accelerometer $\widehat{s_A}$ sequences. The Generator learns the human features of the swipe gestures and generate human-like ones from Gaussian Noise and human sequences $s_T$, $s_A$.

The accelerometer signals are synthesized as random sequences generated from a Gaussian distribution with mean and standard deviation estimated from real accelerometer signals from HuMIdb.

### 4.2.2 Method 2: GAN Synthesis

For this approach, we employ a GAN (Generative Adversarial Network) architecture firstly proposed by Goodfellow *et al.* [40], in which two neuronal networks, commonly named Generator and Discriminator, are trained in adversarial mode. The Generator tries to fool the Discriminator by generating fake samples (touch trajectories and accelerometer signals in this work) very similar to the real ones, while the Discriminator has to discriminate between the real samples and the fake ones created (see Fig. 3 for the details). Once the Generator is trained, then we can use it to synthesize swipe trajectories very similar to the real ones.

The topology employed in both Generator and Discriminator consist of two LSTM (Long Short-Term Memory) layers followed by a dense layer, very similar to a recurrent auto-encoder. The LSTM layers learn the time relationships of human swipe sequences, while the dense layer is used as a classification layer to distinguish between fake and real swipe trajectories in the Discriminator or to build synthetic swipe trajectories in the Generator. To synthesize accelerometer signals, we follow the same GAN architecture described before, but extending the input of the generator from {**x**, **y**} swipe coordinates to {**x**, **y**, **z**} accelerometer axes.

### 4.3 Experimental Protocol

Both GAN networks were trained using more than 10K human samples extracted from the HuMIdb. Training details: learning rate $\alpha = 2 \cdot 10^{-4}$, Adam optimizer with $\beta_1 = 0.5$, $\beta_2 = 0.999$, and $\varepsilon = 10^{-8}$. The system was trained for 50 epochs with a batch size of 128 samples for both Generator and Discriminator. The loss function was '*binary crossentropy*' for the Discriminator and '*mean square error*' for the Generator. The model was trained and tested in *Keras-Tensorflow*.

We generated 12K synthetic samples according to the two methods proposed (up to 18K samples between all groups: 6K human samples, 6K GAN synthetic samples, and 6K handcrafted synthetic samples). Once we have extracted the global features from human and synthetic swipe trajectories and accelerometer data we classify them employing three classification algorithms: an SVM (Support Vector Machine) with an RBF (Radial Basis Function), KNN (K-Nearest Neighbors) with $K = 10$, and RF (Random Forest). The experiments are divided into two different scenarios depending on the synthetic data (i.e. handcrafted or GAN) employed in training: multiclass or agnostic. In multiclass classification, we train and test the classifiers with the same kind of synthetic samples in order to analyze whether the classifier can find discriminative features between both human and bots samples. In the agnostic classification, we train the classifiers using samples of one bot generation method and test with the other one, in order to study whether the classifiers are able to detect bot samples from unknown bot generation methods not seen during the training phase.

In both classification setups, there is no overlap between the data used for training and evaluation. We use 70% of all samples (randomly chosen) as the training set, which is further divided into development (90%) and validation set (10%) in order to choose the best hyper-parameters of the classifiers. The remaining 30% of the samples is used for the evaluation of the system. Both development and evaluation sets are balanced with same number of human and bot samples in each set. All experiments were repeated 5 times (with random selection of the data sets) and the results were computed as the average of the 5 iterations with a standard deviation of $\sigma \sim 0.1\%$.

|  | | Bot Detection | | | | | | | | | | | | | |
|---|---|---|---|---|---|---|---|---|---|---|---|---|---|---|---|
|  | | HandCrafted | | | | | GAN | | | | | HandCrafted+GAN | | | | |
|  | Classifiers | AUC | Acc | Re | Pre | F1 | AUC | Acc | Re | Pre | F1 | AUC | Acc | Re | Pre | F1 |
| Touch | SVM (M) | 99.2 | 94.2 | 89.4 | 98.8 | 93.9 | 98.6 | 95.5 | **95.0** | 95.6 | 95.5 | 93.6 | 85.8 | 82.9 | 88.1 | 85.4 |
| | KNN (M) | 88.3 | 80.6 | 74.7 | 84.8 | 79.4 | 98.6 | 94.6 | 92.0 | 97.0 | 94.5 | 90.0 | 80.1 | 78.0 | 82.3 | 80.0 |
| | RF (M) | **100.0** | **99.9** | **99.9** | **100.0** | **99.9** | 99.3 | 97.3 | 94.7 | **98.2** | 96.4 | 99.7 | 96.5 | 96.8 | 97.7 | 97.3 |
| | SVM (A) | **61.3** | 51.7 | 96.6 | **49.1** | 65.2 | 70.4 | 56.6 | 88.5 | 43.0 | 61.4 | - | - | - | - | - |
| | KNN (A) | 57.5 | **53.8** | 91.9 | 48.3 | 63.3 | 76.7 | 63.6 | 74.5 | **57.0** | 54.6 | - | - | - | - | - |
| | RF (A) | 56.6 | 52.2 | 93.9 | 48.8 | 64.3 | 50.8 | 50.1 | **99.9** | 50.1 | 66.6 | - | - | - | - | - |
| Touch + Acce | SVM (M) | 99.9 | 99.2 | 99.2 | 99.3 | 99.2 | 99.1 | **99.8** | 99.4 | 99.2 | 99.5 | 99.2 | 99.2 | 99.6 | 98.8 | 99.2 |
| | KNN (M) | 99.8 | 99.0 | 98.9 | 99.2 | 99.0 | 98.7 | 99.7 | 99.1 | 99.3 | 99.4 | 99.1 | 98.9 | 99.2 | 98.5 | 98.9 |
| | RF (M) | **100.0** | **99.9** | 99.8 | 99.9 | **99.9** | 99.7 | 99.6 | **99.9** | **99.8** | **99.8** | **99.9** | **99.8** | 99.7 | **100** | **99.8** |
| | SVM (A) | **93.6** | 82.4 | 99.9 | 74.0 | 85.0 | 88.6 | 68.8 | 98.8 | **62.0** | 76.2 | - | - | - | - | - |
| | KNN (A) | 87.7 | **86.0** | 99.9 | **78.2** | **87.7** | 81.2 | 60.1 | 98.6 | 60.0 | 64.7 | - | - | - | - | - |
| | RF (A) | 92.4 | 85.8 | **99.9** | 78.0 | 87.6 | **99.2** | 54.4 | **99.8** | 53.2 | 66.9 | - | - | - | - | - |

Table 4: Bot detection performance metrics in % (AUC= Area Under the Curve, Acc = Accuracy, Re = Recall, Pre = Precision, and F1) for the different scenarios: Multiclass (M), Agnostic (A).

## 5. Results and Discussion

### 5.1 Performance of Bot Detection: Multiclass vs Agnostic Training

Table 4 shows the bot detection performance metrics (%) for different synthetic trajectories (columns) generated when comparing with the human ones. For this experiment the number of training samples (for both human and synthetic samples) is set to $M = 1000$. The results are presented in terms of AUC (Area Under the Curve), Accuracy, Precision, Recall, and F1.

First, we observe that the results achieved for the agnostic classification are always significantly worse (lower performance) than those achieved in multiclass classification as expected. The synthetic samples generated by the two methods present their own specific features and the inclusion of both types of samples in training clearly improves the detection accuracy.

Secondly, when comparing among classifiers we can observe that the RF classifier performs better in multiclass classification meanwhile in agnostic classification, RF is outperformed by KNN.

Finally, we can observe that classifiers trained with both accelerometer and touch samples perform better than those systems trained only with the touch data, especially in agnostic classification, where the multimodal systems doubled their performance. These results suggest the potential of multimodal approaches, even in this challenging scenario where the synthetic training samples are not generated with the same method employed for the evaluation, in which the systems can maintain bot detection rates over 90%.

To better understand the results, Fig. 4 shows the probability functions of the six features proposed for the three types of touch signals (i.e. humans and both synthetic generation methods). Synthetic distributions do not completely fit the human distributions, but they present a behavior like the human samples. First, we can observe that the Move Efficiency of the handcrafted trajectories is equal to 1, this happens because in swipe trajectories with straight line shape the distance and displacement are equal. This is the reason why the multiclass classifiers detect these synthetic trajectories so easily. Note that the Duration (length) of both handcrafted and GAN synthetic swipes were computed as a Gaussian distribution with the same mean and standard deviation as the human ones so both probability distributions are equal. Regarding Distance and Displacement, the GAN trajectories fit worse than the handcrafted ones. We suggest that the main reason for this is that the GAN network generates smoother swipe trajectories than the human ones without abrupt direction changes, causing longer displacements in less distance (like a parabolic function). Finally, the Velocity Profile of both synthetic swipe trajectories are very similar to the human ones, the initial acceleration applied to the function-based trajectories reproduces human behaviors with great similarity while the GAN network learns very realistic Velocity Profiles of human swipe trajectories as well.

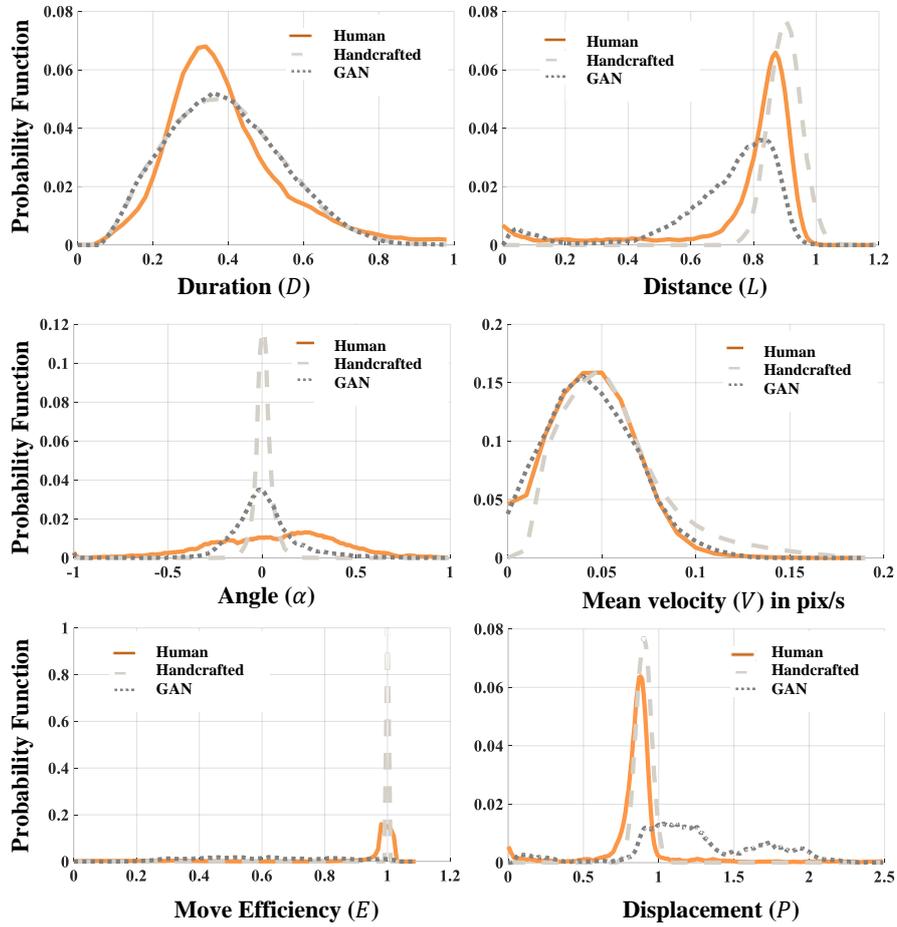

Figure 4: Probability functions of the six global features for Human, Handcrafted, and GAN touch trajectories.

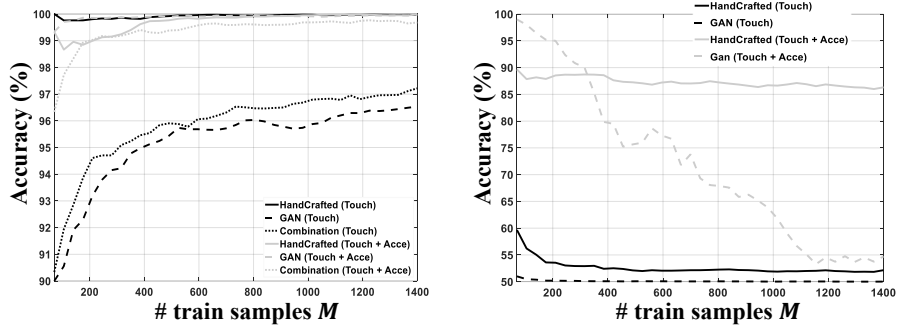

Figure 5: Accuracy curves (%) against the number of train samples ($70 \leq M \leq 1400$) to train the different classifiers in multiclass (left) and agnostic (right) classification scenarios.

## 5.2 Ablation Study: Number of Training Samples

In Fig. 5 we first explore to what extent the number of training samples affects the classification performance. For this, we plot accuracy curves for the best classifier (i.e. RF for multiclass and KNN for agnostic classification) against the number of samples employed to train them ($M$). Remember that both training and evaluation sets are balanced so the number of human ($M_h$) and synthetic ($M_s$) train samples are equal, i.e.: $M_s = M_h = M/2$.

We can observe in Fig. 5 (left) that the accuracy improves when scaling up the number of train samples as we expected. The accuracy improves significantly up to $M = 1000$. On the other hand, it is surprising that the opposite tendency is observed in agnostic classification (Fig. 5 right), where the accuracy rates decay when scaling up the number of train samples. We suggest that the problem in agnostic classification is that classifiers are better trained to detect a specific synthetic generation method, making more difficult for them to detect synthetic samples generated with other methods as we increase the number of training samples with a specific method (i.e. some kind of overfitting to the specific bot generation method used for training).

|                          | **Bot Detection** |      |                |
|--------------------------|-------------------|------|----------------|
| **SVM Classifiers**      | Handcrafted       | GAN  | HandCrafted+GAN |
| One-class (Touch)        | 62.3              | 54.6 | 57.1           |
| One-class (Touch + Acce) | 89.2              | 79.4 | 80.5           |

Table 5: Accuracy rates (%) in bot detection for the one-class SVM classifiers, where the SVM is trained with only human samples and tested with both synthetic generation methods.

## 5.3 Performance of Bot Detection: One-Class Classification

The previous results encourage us to explore one-class classification scenario, where we train the classifier using only the human samples and test with both human and synthetic samples, in order to study whether the classifier is able to detect bots as abnormal human behavior. For this, we employ a SVM classifier that usually works well in one-class classification and set $M = 1000$. Table 5 shows the accuracy rates (%) for one-class SVM bot detection where rows represent the modality of the human samples (i.e. touch or touch plus accelerometer) employed to train the classifiers and in columns the bot generation method employed in the test. We can observe that synthetic samples generated with GAN can fool the classifier more times than the handcrafted samples as we expected, showing the potential of GAN networks to reproduce human trajectories with a great similarity, making almost impossible for the classifier to discriminate between synthetic GAN trajectories and human ones. The fusion of touch trajectories with accelerometer data improves the accuracy rates by more than 30%. GAN networks are not able to reproduce human accelerometer signals as well as touch trajectories, due to the complexity of the accelerometer signals, suggesting again the potential of multimodal approaches to deal with bot attacks.

## 5.4 Performance of Bot Detection: GAN Discriminator

Besides the comparison among different classifier algorithms, we conduct another experiment in which we employ the GAN Discriminator as the classifier. In this experiment the previous feature extraction plus statistical classifier is replaced by a LSTM network (the Discriminator of the GAN). The fact that the Discriminator was trained with synthetic samples generated by the Generator during GAN training could perform better in the classification task than a neural network trained from scratch. Remember that the GAN Discriminator consists of two LSTM (Long Short-Term Memory) layers followed by a dense layer with a sigmoid activation function to discriminate between bots and humans, so in this experiment we tune the number of neurons of these two layers and train a new GAN network for each Discriminator setup.

|  | GAN Discriminator | Bot Detection | | | | | | | | | |
|---|---|---|---|---|---|---|---|---|---|---|---|
|  |  | HandCrafted | | | | | GAN | | | | |
|  |  | AUC | Acc | Pre | Re | F1 | AUC | Acc | Pre | Re | F1 |
| Touch | LSTM (32/16) | **92.2** | **86.8** | **85.7** | **89.2** | **87.4** | 78.3 | 77.8 | 79.2 | 76.7 | 78.3 |
| Touch | LSTM (16/8) | 70.0 | 65.2 | 64.3 | 67.3 | 66.3 | 54.4 | 52.2 | 54.3 | 55.1 | 54.7 |
| Touch | LSTM (32) | 89.1 | 86.7 | 87.7 | 84.7 | 86.1 | 66.2 | 64.3 | 64.1 | 64.5 | 64.4 |
| Touch | LSTM (16) | 89.9 | 87.4 | 89.9 | 86.6 | 87.2 | 52.5 | 52.2 | 53.3 | 51.9 | 52.6 |
| Touch+Acce | LSTM (32/16) | 85.8 | 77.7 | 74.2 | 80.5 | 77.3 | 63.8 | 59.2 | 60.0 | 62.1 | 61.1 |
| Touch+Acce | LSTM (16/8) | 85.5 | 84.4 | 82.1 | 85.7 | 84.1 | 76.2 | 70.4 | 71.3 | 73.4 | 72.2 |
| Touch+Acce | LSTM (32) | 61.1 | 65.3 | 68.4 | 64.4 | 66.3 | 61.7 | 57.7 | 58.8 | 55.6 | 56.7 |
| Touch+Acce | LSTM (16) | **93.4** | **88.8** | **89.9** | **91.2** | **90.8** | 81.1 | 74.4 | 77.3 | 75.5 | 76.4 |

Table 6: Performance metrics in % (AUC= Area Under the Curve, Acc, Pre, Re, and F1) for the different setups of GAN Discriminator in bot detection. In brackets the number of neurons for the first/second LSTM layer respectively used in the Discriminator.

Table 6 shows the bot detection performance metrics (%) for the different synthetic trajectories (columns) generated when comparing with the human ones. In rows, the different GAN Discriminator setups chosen for this experiment: two LSTM layers with 32 and 16 neurons respectively, two layers with 16 and 8 neurons respectively, one layer with 32 neurons, and one layer with 16 neurons.

First, it is surprising to observe that the GAN Discriminator performs better detecting synthetic handcrafted samples, even when the GAN Discriminator was trained only to discriminate between GAN synthetic and human samples. According to these results the GAN Discriminator can perform better than statistical classification algorithms as abnormal human behavior detector (e.g. agnostic and one-class classification scenarios).

| Method | Cognitive | Behavioral | Usability | Attack Protection |
|---|---|---|---|---|
| Audio CAPTCHA | *** | * | * | * |
| Image CAPTCHA | *** | * | * | * |
| Text CAPTCHA | *** | * | * | * |
| reCAPTCHA v3 | * | ** | *** | * |
| Touch CAPTCHA [41] | ** | *** | *** | Unknown |
| Gesture CAPTCHA [42] | ** | *** | ** | *** |
| **BeCAPTCHA (Ours)** | ** | *** | *** | *** |

Table 7: Characteristics of several CAPTCHA methods. We rate each factor as low (*), medium (**) and high (***).

Regarding GAN Discriminator setups, configurations with larger number of neurons (i.e. 32 neurons in the first layer and 16 in the second one) seem to perform better for touch trajectories, and the opposite for the fusion with the accelerometer signals. We suggest that smooth and complex signals such us touch gestures need larger GAN Discriminator setups to be detected meanwhile more simple and noisy signals such us the accelerometer ones can be detected with smaller GAN Discriminator setups.

## 6. Comparison and Complementarity with other CAPTCHAs

Table 7 shows some of the main features of different existing CAPTCHA methods. Audio, image, and text-based CAPTCHAs have been defeated by machine learning algorithms. As an example, in [20] the authors designed an AI-based system called UnCAPTCHA to break Google's most challenging audio reCAPTCHAs. The text-based CAPCTHA was defeated by Bursztein *et al.* [19] with 98% accuracy using a ML-based system to segment and recognize the t ext. Finally, the last version of the Google CAPTCHA, named reCAPTCHAv3, is transparent for the user and measures mouse dynamics and web browsing interactions between the user and the web site to decide whether the user is a bot or not. This version was recently hacked in [21] by synthetizing mouse trajectories using reinforcement learning techniques. The main problem of these CAPTCHA methods is that they only measure cognitive human skills (e.g. character recognition from distorted images, class-objects identification in a set of images or speech translation from distorted audios). Trying to ensure a very accurate bot detection makes these CAPTCHAs difficult to perform even for humans.

The main goal of new generation bot detection algorithms like reCAPTCHA v3 is to focus more in human behavioral skills rather than cognitive ones, as behavioral skills reveal inner human features useful for bot detection just with simple gestures like swipes. In that line of work exploiting simple natural behaviors instead of complex cognitive challenges, two works closely related to our proposed Be CAPTCHA are [41] and [42].

In [41], the authors developed a gesture-based CAPTCHA for mobile devices in which the participants are asked to move objects over the screen to solve the CAPTCHA. In the paper the authors do not evaluate the performance of the proposed CAPTCHA system to discriminate between human and bots. Their algorithm demonstrated to be more user friendly than other existing methods like Google reCAPTCHA. Their success rate (measured as the percentage of CAPTCHA tests that users completed successfully) is 100% versus 91% achieved by the Google reCAPTCHA method. Hupperich et al. [42] proposed a mobile CAPTCHA system based on performing simple gestures (e.g. fishing, hammering, drinking) while holding the mobile device to solve the CAPTCHA. For this, they used the data extracted from the accelerometer and gyroscope and applied machine learning classifiers (Bagging Tress, Random Forest and KNN). Their results achieved over 90% of performance for gesture recognition, demonstrating the suitability of sensor-based CAPTCHA to improve traditional CAPTCHAs in a mobile scenario. These methods were not evaluated when synthetic samples are used to attack the system and their performance analysis were focused on the success rate of humans solving the CAPTCHA challenges.

It is important to highlight that our proposed BeCAPTCHA is compatible with previous CAPTCHA technologies and it could be added as a new cue to improve existing bot detection schemes in a multiple classifier combination [43]. In fact, BeCAPTCHA can be easily extended to consider other inputs beyond the swipe signals considered in our experiments, e.g.: web browsing, texting, solving other CAPTCHAs, etc.

## 6.1 User Perception Survey

At the end of the acquisition, the participants in the HuMIdb database completed a questionnaire about CAPTCHA technologies and ethics issues. The survey includes responses from 600 participants from 14 different countries (see Section 3 for details). The survey included three questions related to the understanding and perception of CAPTCHA technologies by the users:

- Question 1: '*Do you know what is a CAPTCHA system?*' 33% of the participants didn't know, which shows the lack of information about one of the most common malicious malware in mobile devices nowadays.
- Question 2: '*Do you think mobile apps that use biometric user information are privacy invasive?*' most of them (76%) answered affirmative.
- Question 3: '*If this information were useful to improve your security and your confidence in web navigation, would you be willing to share it anonymously?*' 82% of the participant's answers were also affirmative.

According to these results, we can conclude that most mobile users are reluctant to share their biometric information in mobile apps, but they are willing to share it in case they are given a clear benefit in terms of security and confidence.

## 7. Conclusions and Future Work

We introduce a new bot detection system for smartphones based on the analysis of behavioral information from inbuilt sensors: BeCAPTCHA. Results are provided by combining touchscreen and accelerometer data, but the methods are presented in a general way and therefore BeCAPTCHA directly allows incorporating information from additional sensors.

As we discussed in the introduction, the behavioral information acquired through mobile sensors describe inner human features, such as neuromotor abilities, cognitive skills, human routines, and habits. All these patterns can help to develop new bot detection algorithms for mobile scenarios. Our goal in this paper has been to go a step forward on the bot detection field focused on mobile scenarios by implementing CAPTCHA methods that exploit mobile sensor signals during human-mobile interactions. For this, we present a novel multimodal mobile database HuMIdb that comprises 14 mobile sensors captured from 600 users in an unsupervised scenario. Although in this paper we focus on the HuMIdb for bot detection, this new dataset offers many other research opportunities related to modeling and exploiting the human-machine interaction in smartphones.

We have evaluated our proposed BeCAPTCHA approach combining swipe touchscreen trajectories and accelerometer signals extracted from HuMIdb (human samples) vs very realistic synthetic trajectories (bot samples) generated with two methods: GAN deep learning and handcrafted. We provide results in various experimental configurations and classifiers, considering or not synthetic bot data for training BeCAPTCHA (multi-class, agnostic, and one-class). Bot detection results for agnostic classification (i.e. training with one synthetic bot method and testing with the other method) and one-class classification (i.e. training only with the human samples) just using touch gestures are poor with accuracies of around 60%, but the combination with accelerometer data improves the results to the range 80-90% of accuracy. In addition the case of multi-class training (i.e. training with both bot data generation methods) achieves very good performance, with results against very realistic synthetic attacks of over 90% of accuracy for bot detection. Regarding classifiers, Random Forest (RF) seems to perform the best in multi-class scenario while K-Nearest Neighbors (KNN) performs better in the agnostic scenario. In addition, the number of samples (human and bot) employed to train the classifiers affect considerably the performance, meanwhile in multi-class scenario, classifiers perform better as we increase the amount of samples to train them. The opposite tendency is observed in agnostic scenario, where the classifiers reduce their capacity to detect bot samples from other methods as we increase the amount of training data to detect a specific kind of synthetic bot samples. Finally, employing the GAN Discriminator as a classifier reveals the potential of this LSTM network to detect bot samples generated using the handcrafted method, with a performance like using RF in multi-class scenario. Considering that the GAN Discriminator is only trained with human and GAN Generator samples, the potential of the GAN Discriminator for agnostic and one-class classification scenario is patent.

We strongly believe that the combination of these behavioral signals with traditional CAPTCHA methods can harden significantly existing algorithms for bot detection. The expected improvements are even larger when considering additional mobile sensors in extended BeCAPTCHA implementations beyond touchscreen and accelerometer data.

For future works, we will explore the addition of new sensors on top of touchscreen and accelerometer data (as available in our HuMIdb dataset,[2] see Section 3), new approaches that exploit the complementarity between tasks and sensors, and smart fusion to exploit multiple sensors and the heterogeneity of the data [43]. Moreover, we will explore new methods to generate bot samples, the handcrafted method could be combined with more realistic velocity profiles based on human kinematic profiles [44], and different shapes that beter mimic human behaviors.

## Appendix A. Android App and Task Description

Fig. 6 shows all tasks included in the HuMI database. The task *a* is designed to acquire keystroking from fixed and free text. In tasks *b* and *d*, the users have to perform both swipe up and swipe down gestures to complete both tasks, meanwhile the task *c* is focused on tap gestures. Tasks *e* and *f* are designed to draw in the air with the smartphone a circle and a cross respectively. Task *g* records the user saying '*I am not a robot*', and finally, in task *h* the user has to draw with the finger the digits 0 to 9 over the touchscreen. Note that the 14 sensors available (see Section 3) are acquired during the execution of all tasks, although some sensors present a key role in some of them.

---

[2] https://github.com/BiDAlab/HuMIdb

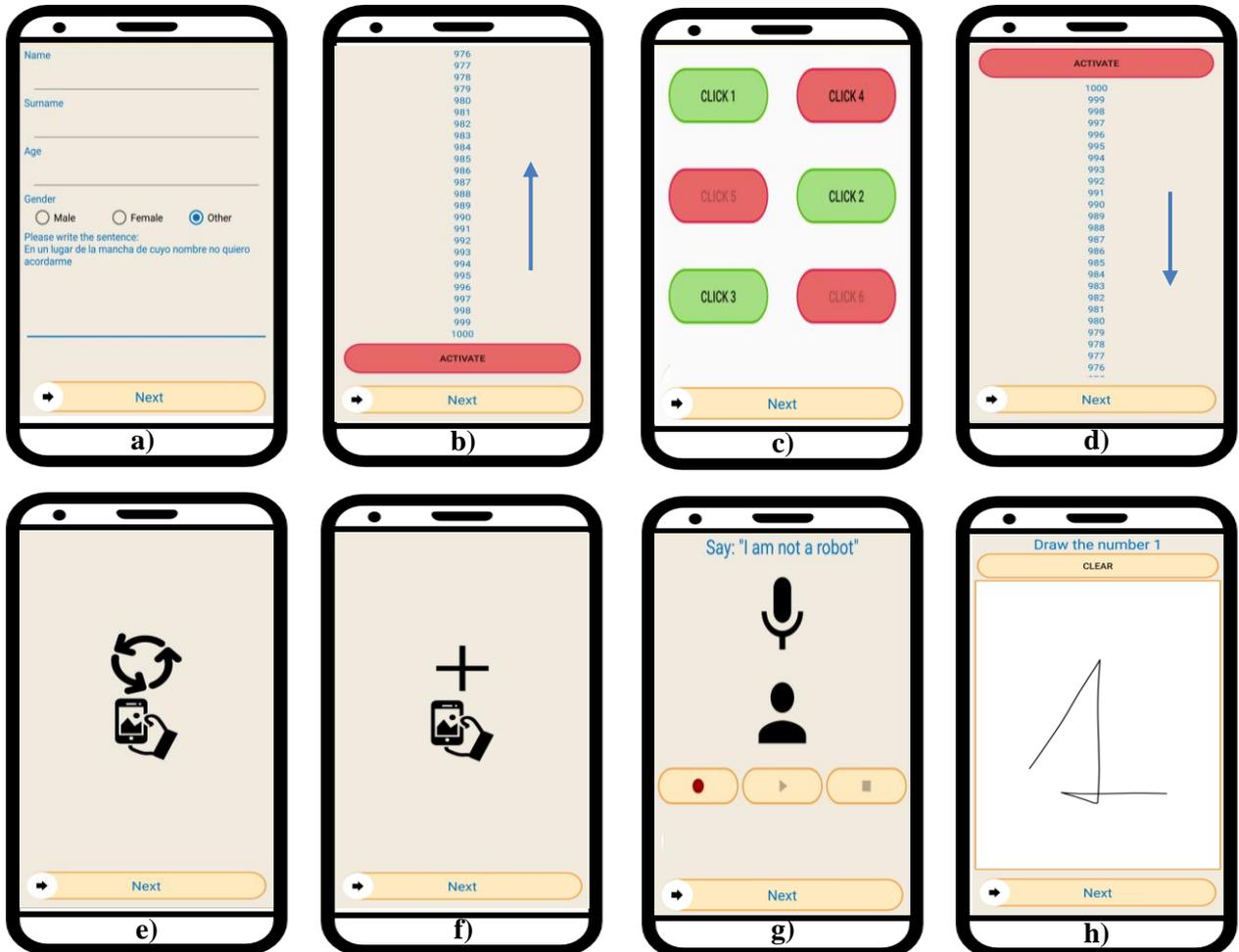

Figure 6: The mobile interfaces designed for the 8 mobile HuMI tasks: *a)* keystroking, *b)* swipe up, *c)* tap and double tap, *d)* swipe down, *e)* circle hand gesture, *f)* cross hand gesture, *g)* voice, and *h)* finger handwriting.

For example, the accelerometer signal is captured during the entire session even though it could be more relevant in tasks *e* and *f*. This heterogeneous information can be used to improve the patterns obtained from the main sensor for each task. Additionally, all tasks have a right swipe button that is acquired in addition to the swipe patterns.

**Acknowledgements**

This work has been supported by projects: PRIMA (H2020-MSCA-ITN-2019-860315), TRESPASS-ETN (H2020-MSCA-ITN-2019-860813), BIBECA (RTI2018-101248-B-I00 MINECO/FEDER), and BioGuard (Ayudas Fundación BBVA a Equipos de Investigación Científica 2017). Spanish Patent Application P202030066.